\documentclass[twocolumn,twocolappendix]{aastex631}
\usepackage{natbib}
\usepackage{amssymb,amsmath}
\usepackage{makecell}
\usepackage{soul}
\bibpunct{(}{)}{;}{a}{}{,}
\begin{document}

\title{A Novel Approach to Identify Unseen Companions: An In-Depth study of GDR3 220\_28}
\shorttitle{An In-Depth study of GDR3 220\_28}

\author{T. B. Cysil}
\affiliation{Department of Physics and Electronics, CHRIST (Deemed to be University), Bangalore, 560029, India}
\author{Blesson Mathew}
\affiliation{Department of Physics and Electronics, CHRIST (Deemed to be University), Bangalore, 560029, India}
\affiliation{Center of Excellence in Astronomy and Astrophysics, Department of Physics and Electronics, CHRIST (Deemed to be University), Bangalore, 560029, India}
\author{B. Shridharan}
\affiliation{Tata Institute of Fundamental Research, Mumbai 400005, India}
\author{S. Muneer}
\affiliation{Indian Institute of Astrophysics, Bangalore, 560034, India}
\author{Pramod Kumar S}
\affiliation{Indian Institute of Astrophysics, Bangalore, 560034, India}
\author{S. Nidhi}
\affiliation{Department of Physics, Maharaja Ranjit Singh Punjab Technical University, Bathinda, 151001, India}
\author{Sreeja S Kartha}
\affiliation{Department of Physics and Electronics, CHRIST (Deemed to be University), Bangalore, 560029, India}
\affiliation{Center of Excellence in Astronomy and Astrophysics, Department of Physics and Electronics, CHRIST (Deemed to be University), Bangalore, 560029, India}
\author{Suman Bhattacharyya}
\affiliation{Department of Physics and Electronics, CHRIST (Deemed to be University), Bangalore, 560029, India}
\author{T. Robin}
\affiliation{Centre for Space Research, North-West University, Potchefstroom, 2520, South Africa}

\correspondingauthor{T. B. Cysil \\ ORCID: \href{https://orcid.org/0000-0001-6632-1139}{0000-0001-6632-1139}}
\email{cysilbabytom@gmail.com}
\shortauthors{T. B. Cysil et al.}

\begin{abstract}
Emission lines in the spectrum and the excess seen in the spectral energy distribution (SED) at infrared (IR) wavelengths serve as crucial indicators of accretion activity and the presence of gas and dust in the circumstellar environment of a young star. As accretion ceases over several million years, the H$\alpha$ emission diminishes. However, IR excess combined with negligible H$\alpha$ emission could also extend to indicate certain close binary systems, where a circumbinary disk contributes to the observed excess. Additionally, in unresolved binaries, companion stars can contribute to the photometry, leading to inaccuracies in parameter measurements and skewed positions in the color-magnitude diagram (CMD). We identify a photometrically peculiar object from a survey of the Cepheus star-forming regions (SFRs), Gaia DR3 2205212960562531328, exhibiting a shift from its expected position in the CMD, as indicated by the Gaia optical photometry. This peculiarity drives our research, prompting an investigation through an approach that combines astrometric accuracy, photometric modeling and spectroscopic studies. The Gaia XP spectrum of the source shows a systematic blue excess, indicating an additional hotter unresolved component. Two-component SED modelling yields a companion with $T_{\rm eff}\sim27000$ K and $\log g\ge4.5$, consistent with sdB stars. Further CMD and spectroscopic analyses support a binary system consisting of an A0-type PMS star with an sdB companion. In addition, ZTF photometry shows tentative evidence for a periodicity of 0.498 days. This study paves the way for a systematic search of similar binary systems that would otherwise remain unidentified.
\end{abstract}


\section{Introduction}

The evolution of pre-main sequence (PMS) stars is strongly influenced by the accretion processes from its circumstellar environment \citep{Hayashi, hartmann, 1998AJ....116.1816H, 2010A&A...510A..72F}. An important observational trait of PMS stars is the presence of various emission lines in their spectra \citep{1960ApJS....4..337H, 1979ApJS...41..743C, 2004AJ....127.1682H}, which are known to arise from accretion columns, inner disk and outflows \citep{2004ApJ...617..406M, 2016ARA&A..54..135H, 2016ApJ...831..169S, 2018A&A...609A..87N, 2023A&A...671A.129T}. Among these emission lines, the H$\alpha$ emission line has been long used to estimate the mass accretion rate in PMS stars \citep{2004AJ....128.1294C, 2006MNRAS.370..580K, 2010ApJ...715....1D, 2012MNRAS.421...78S, 2015MNRAS.452.2837M, 2017A&A...600A..20A, 2017A&A...602A..33F}. In addition, the spectral energy distribution (SED) of a PMS star is an observable feature that characterizes the presence of a circumstellar environment, such as envelopes or disks, surrounding the star. This environment influences the SED, particularly at infrared (IR) ($\lambda>1 \mu m$) wavelengths, contributing to the excesses observed in many PMS stars \citep{kenyon1987, 2004A&A...426..151A, 2009ApJ...699.1067M}. Over the period of a few Myrs, the accretion processes gradually decline and the H$\alpha$ emission gradually disappears \citep{1998ApJ...495..385H, 2012A&A...543A..59M, 2015MNRAS.452.2837M, 2019AJ....157..159A}. During this stage of disk evolution, the inner gaseous disk dissipates with potential planet formation occurring in the outer regions of the disk \citep{2023ASPC..534..539M}; with the presence of dust around the star contributing to the observed excesses at IR wavelengths.

From a different perspective, the presence of IR excess, lacking substantial H$\alpha$ emission, could also result from close binary systems, where the IR excess is likely attributed to a circumbinary disk \citep{van2019binary, kluska2022population}. Moreover, in the specific case of unresolved binaries, a companion star may substantially contaminate the light attributed to the primary, thereby leading to incorrect stellar and disk parameter measurements \citep{2000prpl.conf..703M}. This may lead to skewed positions in their color-magnitude diagram (CMD), with specific systems populating regions toward the left/bottom of the zero age main sequence (ZAMS). This is seen in unresolved binary systems, such as those between white dwarfs and main sequence stars (WD+MS Binaries; \citealp{2021MNRAS.506.5201R}) and in hot subdwarf (sdB/O) stars. sdB/O stars are evolved stellar objects representing the `extreme' horizontal branch phase of low-mass stellar evolution \citep{2009ARA&A..47..211H, 2016PASP..128h2001H}. They form primarily via binary interactions, although single‐star envelope ejection channels have also been shown to contribute \citep{2002MNRAS.336..449H, 2003MNRAS.341..669H}, and are known to exhibit binarity of $\sim40-50\%$ \citep{2015MNRAS.450.3514K, 2022A&A...661A.113G}. In cases where binary systems remain undetected through traditional observational techniques, precise photometric shifts in the CMD offer a promising alternative for identification. This approach could be particularly useful for addressing the observed under-representation of these systems compared to theoretical predictions from stellar population models \citep{2003MNRAS.341..669H, 2008ApJS..174..223B, 2017ApJS..230...15M}.

Towards the identification of such systems, we compile data from an extensive survey of the Cepheus star-forming regions (Cysil et al. submitted), identifying objects that exhibit IR excess and negligible H$\alpha$ emission. On further investigation, we observe a shift in the position of one of these objects in its optical CMD, towards the left of the ZAMS, suggesting possible contamination in its optical photometry from a companion star. The observed characteristics of this source motivate the focus of this work. The extensive analysis presented herein serves as a guide for future studies on similar classes of objects. This work is structured as follows: Section \ref{data} details the process of the identification of this source along with the various data products and services used in this work. Section \ref{analysis} describes this source by its astrometric, photometric and spectroscopic properties. It also discusses, in detail, the peculiar nature of the source. A summary of the analyses conducted in this work is presented in Section \ref{summary}.

\section{Data used in this work}\label{data}

This work utilizes data queried from the European Space Agency mission Gaia. Launched in 2013, the operation of Gaia involves measuring the parallax and proper motions of stars with unmatched accuracy. We make use of Gaia DR3 \citep{2023A&A...674A...1G}, to query the prominent star-forming regions (SFRs) of the Cepheus complex \citep{2008hsf1.book..136K}. The Cepheus SFRs offer a diverse range of environments and conditions. This includes various cloud complexes and OB associations that host PMS stars in various stages in their formation and evolution, providing an intriguing opportunity to discover and investigate unconventional objects.

\subsection{Sample Selection}

The initial query to the Cepheus SFRs from Gaia DR3 is reduced to contain only those objects that follow the astrometric constraints given below,

\begin{itemize}
    \item $\pi / \sigma_{\pi} \geq 3$,
    \item $\pi \geq 0.5$,
    \item Renormalised Unit Weight Error (RUWE) $\leq 1.4$
\end{itemize}

where $\pi$ is the parallax and $\sigma_{\pi}$, the uncertainty in parallax. Following these constraints, the list of objects is further cross-matched with 2MASS \citep{skrutskie2006two} and AllWISE \citep{cutri2013explanatory, cutri2021vizier} databases in order to study the IR signature of these objects. The cross-match was performed adopting a matching radius of 1\arcsec, favoring the nearest neighbor approach. Any source lacking a counterpart within the matching radius of 1\arcsec\ was discarded. Quality cuts in the WISE photometric bands are used to filter out bad data, following the method given in \cite{koenig2014classification}. Specifically, we require a non-null photometric uncertainty, $\texttt{w?sigmpro}$, in the corresponding WISE band, together with cutoffs $\texttt{w1rchi2}<(\texttt{w1snr}-3)/7$ for the WISE \textit{W1} band; $\texttt{w2rchi2}<0.1\times \texttt{w2snr}-0.3$ for \textit{W2}; $\texttt{w3rchi2}<0.125\times \texttt{w3snr}-1$ for \textit{W3}; and $\texttt{w4rchi2}<0.2\times \texttt{w4snr}-2$ for \textit{W4}. The distance of the sources are obtained from \cite{2021AJ....161..147B}. We queried the 3D dust reddening to these objects from \cite{green}, which is then converted to extinction in the visual band $A_V$, assuming the optical total-to-selective extinction ratio, $R_V=3.1$ \citep{1989ApJ...345..245C}. This process returned a list of 50,953 objects, containing astrometric and photometric data from the Gaia, 2MASS, and AllWISE databases, along with corresponding distance measurements and extinction values.

\begin{figure}
\begin{center}
	\includegraphics[width=\columnwidth]{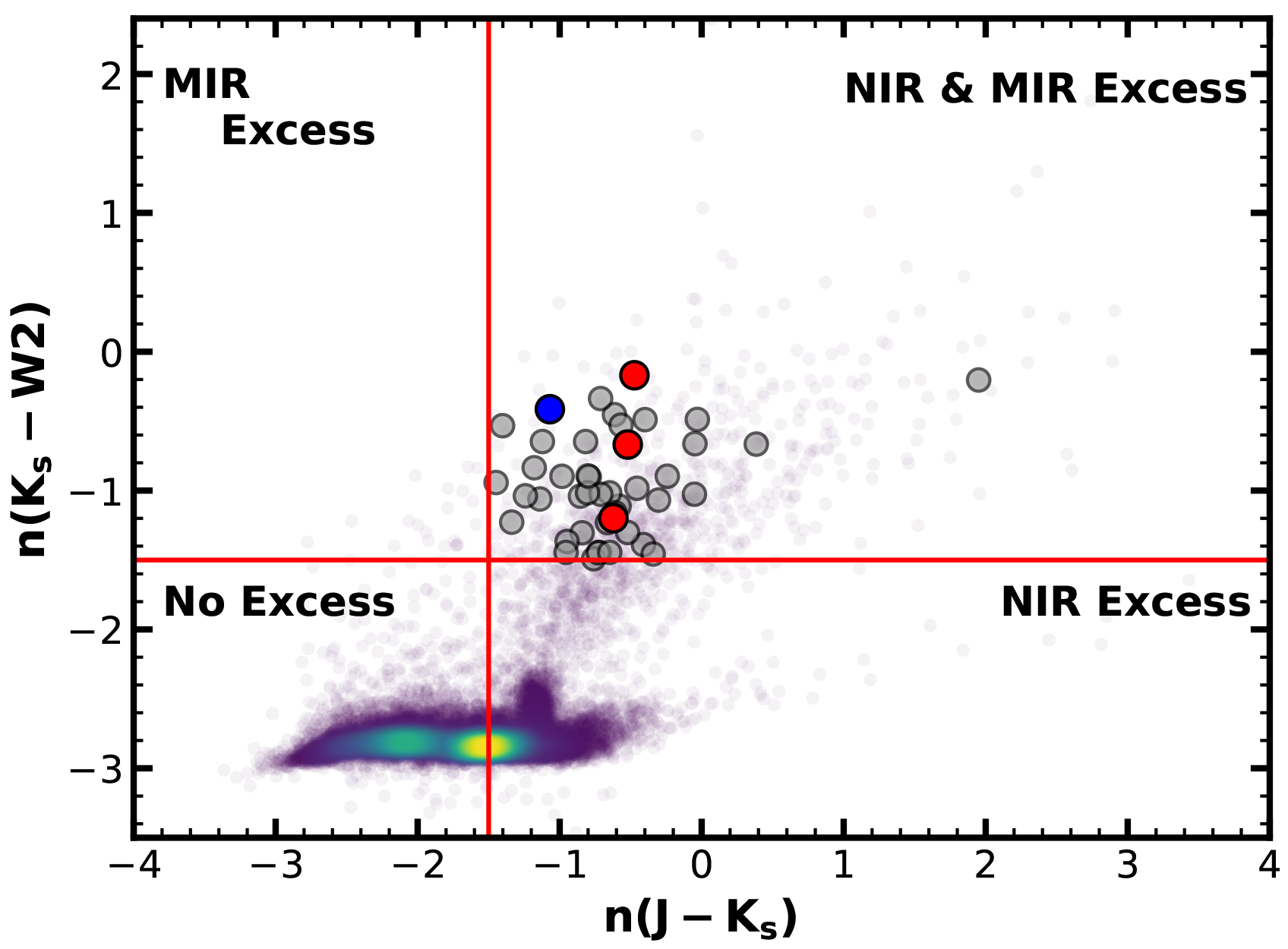}
    \caption{The $n_{K_s-W2}$ index plotted against $n_{J-K_s}$ for the sample of 50,953 objects (background density scatter shown in {\it viridis}). The grey circles represent sources with pseudo EW $>$ 0 and the red circles represent objects with the corrected EW $>$ 0. The blue circle represents the object studied in this work, GDR3 220\_28.}
    \label{Sel}
\end{center}
\end{figure}

This sample is then reduced to contain only Gaia detections which show IR excess, as well as the H$\alpha$ spectral line in absorption. The color indices\footnote{The color index;\quad \large{$n_{\lambda_1-\lambda_2}=\frac{log(\frac{\lambda_2F_{\lambda_2}}{\lambda_1F_{\lambda_1}})}{log(\frac{\lambda_2}{\lambda_1})}$}}, $\mathrm{n_{K_s-W2}}$ and $\mathrm{n_{J-K_s}}$ are used to classify the presence of mid-IR and near-IR excess, respectively. A cutoff is applied, with $\mathrm{n_{K_s-W2}>-1.5}$ indicating mid-IR excess and $\mathrm{n_{J-K_s}>-1.5}$ indicating near-IR excess \citep{1993ApJ...406..122A, 2021RAA....21..288S}. The constraints on H$\alpha$ emission strength is made using the H$\alpha$ pseudo equivalent width (EW) measurements given in Gaia DR3 \citep{2023A&A...674A..28F}. 

The sample is thus reduced to 45 detections which clearly show excess in the near-IR and mid-IR wavelengths, and have pseudo EW of H$\alpha$ greater than 0 (i.e, H$\alpha$  is in absorption). These objects are further refined by converting the pseudo EW measurements of Gaia into EW using a conversion given in \cite{2022A&A...668A.156S} and retaining only those objects which have this corrected EW measurements greater than 0. This reduces the sample to 4 objects. The brightest of these 4 objects, Gaia DR3 2205212960562531328, having a visual \textit{V} band magnitude $\sim13$, also exhibits a shift in its Gaia CMD, suggesting contamination of its optical photometry, which will be introduced in Section \ref{cmdsec} and discussed in more detail in Section \ref{binary}. This object (hereafter referred to as `GDR3 220\_28' or simply `source' when already in context) is taken for further analysis. Figure \ref{Sel} shows the distribution of this sample of 50,953 objects along the various cutoffs discussed above in the $n_{K_s-W2}$ versus $n_{J-K_s}$ index space.

\subsection{Spectra of GDR3 220\_28}

\begin{figure*}
\begin{center}
	\includegraphics[width=2\columnwidth]{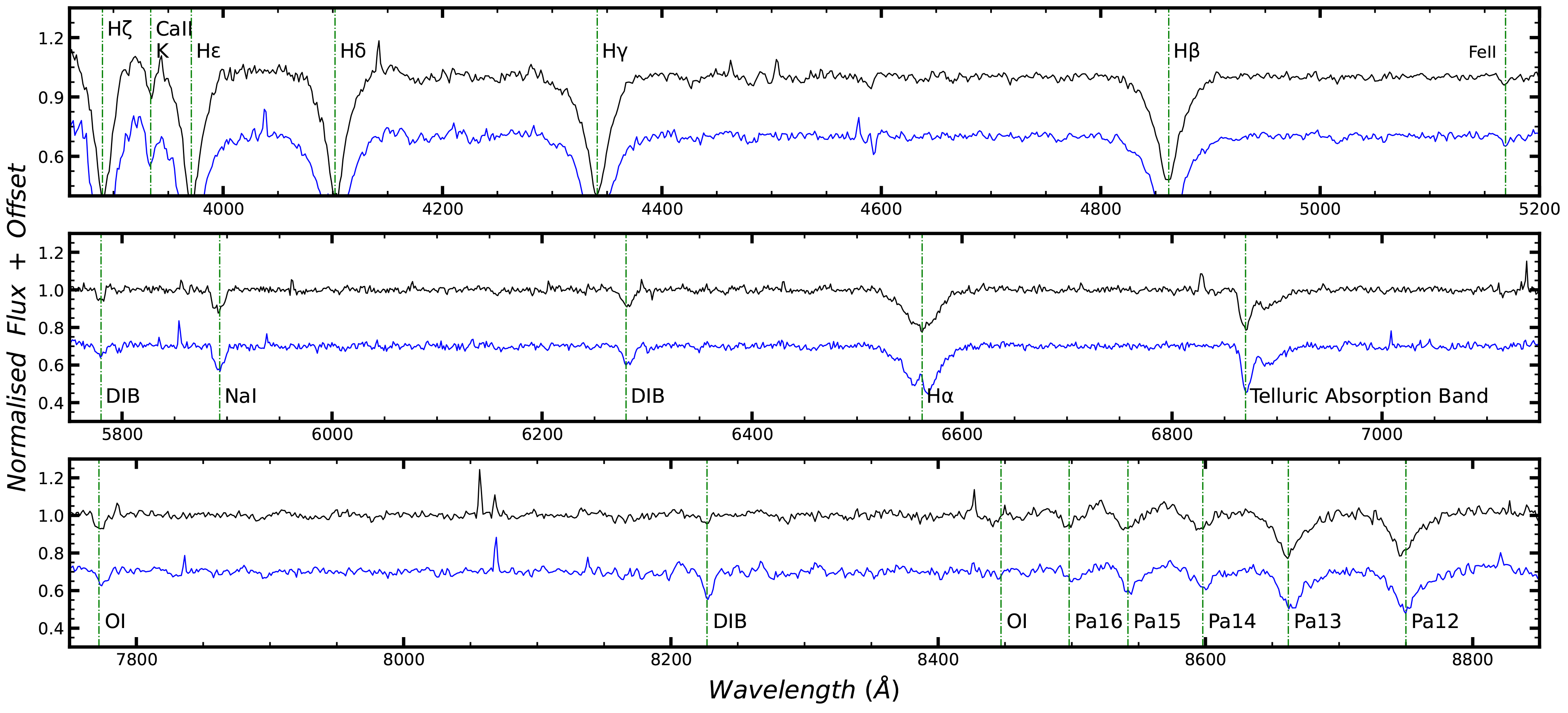}
    \caption{Spectra of GDR3 220\_28 taken at HCT, showing major Hydrogen lines. The spectra in black and blue are observed on 2023 January 21 and 2023 January 31, respectively. A few important lines are labeled.}
    \label{Spectra}
\end{center}
\end{figure*}

The spectra of GDR3 220\_28 are observed using the Hanle Faint Object Spectrograph Camera (HFOSC), mounted on the 2-m Himalayan Chandra Telescope (HCT), over multiple epochs. HFOSC covers a wavelength range of 3500 to 9000 \AA\ using a set of 11 grisms. The spectra obtained for this work are observed using grism 7 (Gr7: 3800 - 6840 \AA; R$\sim$1200) and grism 8 (Gr8: 5800 - 8350 \AA; R$\sim$2000) with the 167l slit. The choice of grisms used for observations are to facilitate the observation of the H$\alpha$ spectral line. A total of four spectra are observed -- on 2023 January 16 (epoch 1), 2023 January 21 (epoch 2), 2023 January 31 (epoch 3), and 2023 February 08 (epoch 4), respectively. All spectra were obtained with an exposure time of 900~s. The median continuum S/N measured around 6000~\AA\ is approximately 134 for spectra observed using Gr7 and 93 for spectra observed using Gr8. Figure \ref{Spectra} displays the spectra observed for GDR3 220\_28 at epochs 2 and 3. In addition to the HCT observations, we also analyse the low-resolution, flux-calibrated BP/RP (XP) sampled spectrum of GDR3 220\_28 available in Gaia DR3 \citep{2023A&A...674A...1G, 2023A&A...674A...2D, 2023A&A...674A...3M}.

\section{Analysis and Results}\label{analysis}

This section provides a detailed analysis of GDR3 220\_28, examining its astrometric and photometric parameters, as well as spectroscopic variability, and proceeds to discuss the results.

{\renewcommand{\arraystretch}{1.2}
\begin{table}[h]
    \centering
    \caption{Astrometric parameters of GDR3 220\_28.}
    \begin{tabular}{|c|c|}
    \hline
         Astrometric Parameter & Value\\
         (Gaia DR3)&\\
         \hline
         RA & 337.21098542 $deg$ $\pm$ 0.01 $mas$\\
         &(22$^{h}$ 28$^{m}$ 50$^{s}$.64)\\
         \hline
         Dec & 63.27173039 $deg$ $\pm$ 0.01 $mas$\\
         &(63\textdegree 16\arcmin 18\arcsec.23)\\
         \hline
         PMRA & -1.02 $\pm$ 0.01 $mas\ yr^{-1}$\\
         \hline
         PMDec & -2.07 $\pm$ 0.01 $mas\ yr^{-1}$\\
         \hline
         Parallax & 1.127 $\pm$ 0.012 $mas$\\
         \hline
         Distance & 872.5$^{+8.4}_{-8.9}\ pc$\\
         \citet{2021AJ....161..147B}&\\
         \hline
         RUWE & 1.1015\\
         \hline
    \end{tabular}
    \label{astpar}
\end{table}}

\begin{figure}
\begin{center}
	\includegraphics[width=\columnwidth]{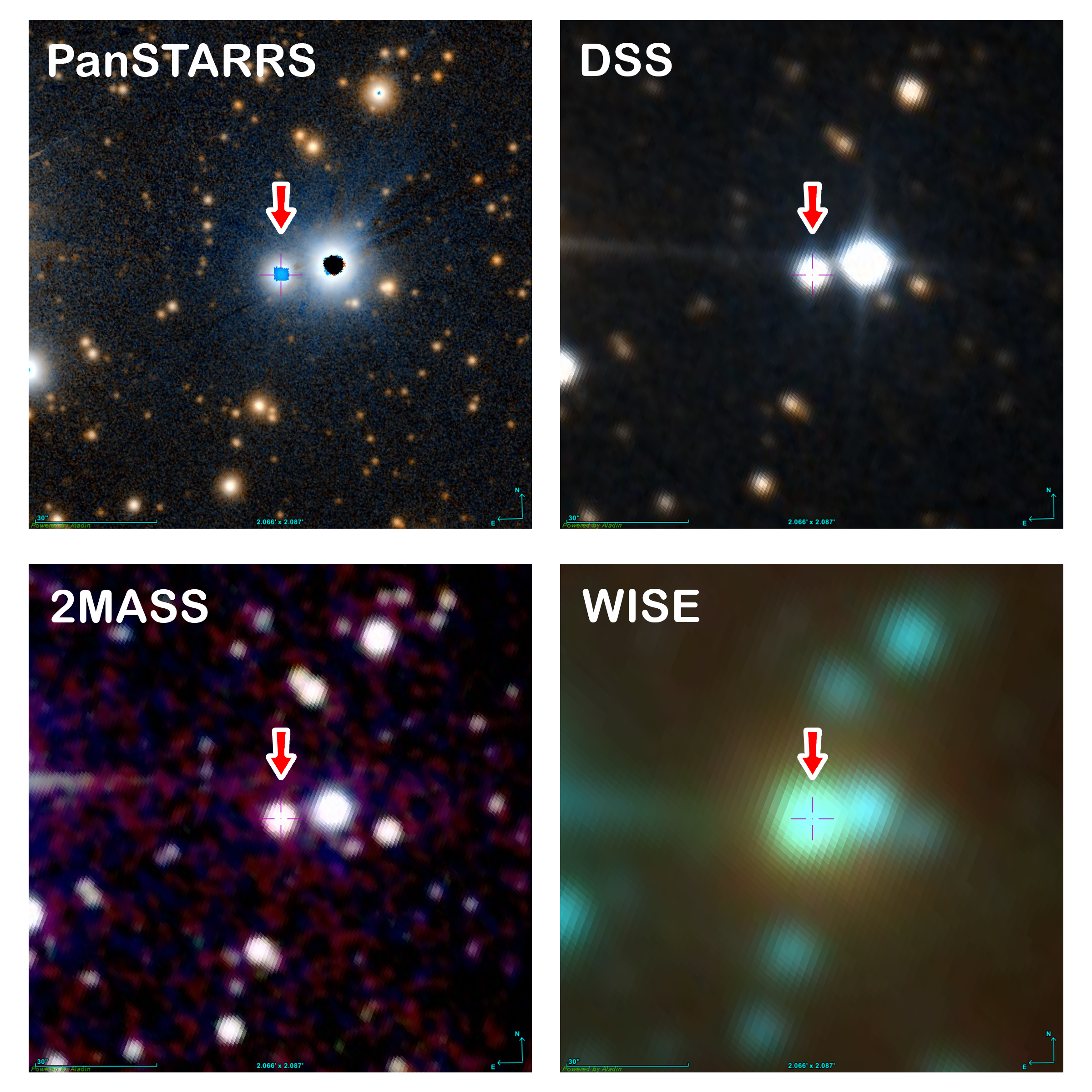}
    \caption{Figure shows the color composite images (2\arcmin x2\arcmin) of GDR3 220\_28 (marked by the red arrow) from the PanSTARRS (top left), DSS (top right), 2MASS (bottom left), and WISE (bottom right) surveys.}
    \label{Aladin}
\end{center}
\end{figure}

Table \ref{astpar} lists some of the astrometric parameters of the source obtained from Gaia DR3. Figure \ref{Aladin} shows GDR3 220\_28 in optical and IR images distributed over 4 panels. The PanSTARRS DR1 composite z-g image \citep{2018AAS...23110201C}, the DSS\footnote{\url{https://dx.doi.org/10.26131/IRSA441}} composite image, the 2MASS composite \textit{J-H-K$_{s}$} image \citep{skrutskie2006two} and the AllWISE composite (Red-\textit{W4}, Green-\textit{W2}, Blue-\textit{W1}) images \citep{cutri2013explanatory} are shown.

\subsection{Photometric Quality Assessment}\label{phot}

Since the peculiarity of GDR3 220\_28 is discernibly dependent on the IR excesses along with the shift in its optical CMD, we run quality checks on the photometry in the IR (2MASS, WISE) and optical (Gaia) wavelengths to ensure the credibility of this observed peculiarity. The photometric quality of the 2MASS  \textit{J}, \textit{H} and \textit{K$_s$} filters for GDR3 220\_28 is tagged with $ph\_qual=AAA$, implying reliable photometry \citep{cutri20032mass}. Quality cuts in the WISE bands as given in \cite{koenig2014classification} are used along with a query for the source in the WISE image archive, to verify the goodness of the WISE photometry. With respect to Gaia photometry, we estimate the corrected BP and RP flux excess factor $C^{*}$, as described in \cite{riello2021gaia}, to ensure no contamination from any nearby sources. The Gaia photometry is well behaved and free from contamination when the value of $C^{*}$ is closer to 0. We estimated $C^{*}$ for GDR3 220\_28, equal to $0.0042$, which falls very well within $1\sigma$ error of the Stetson and Ivezic standards \citep{stetson2000homogeneous, ivezic2007sloan}.

\subsection{Color Magnitude Diagram}\label{cmdsec}

\begin{figure}
\begin{center}
	\includegraphics[width=\columnwidth]{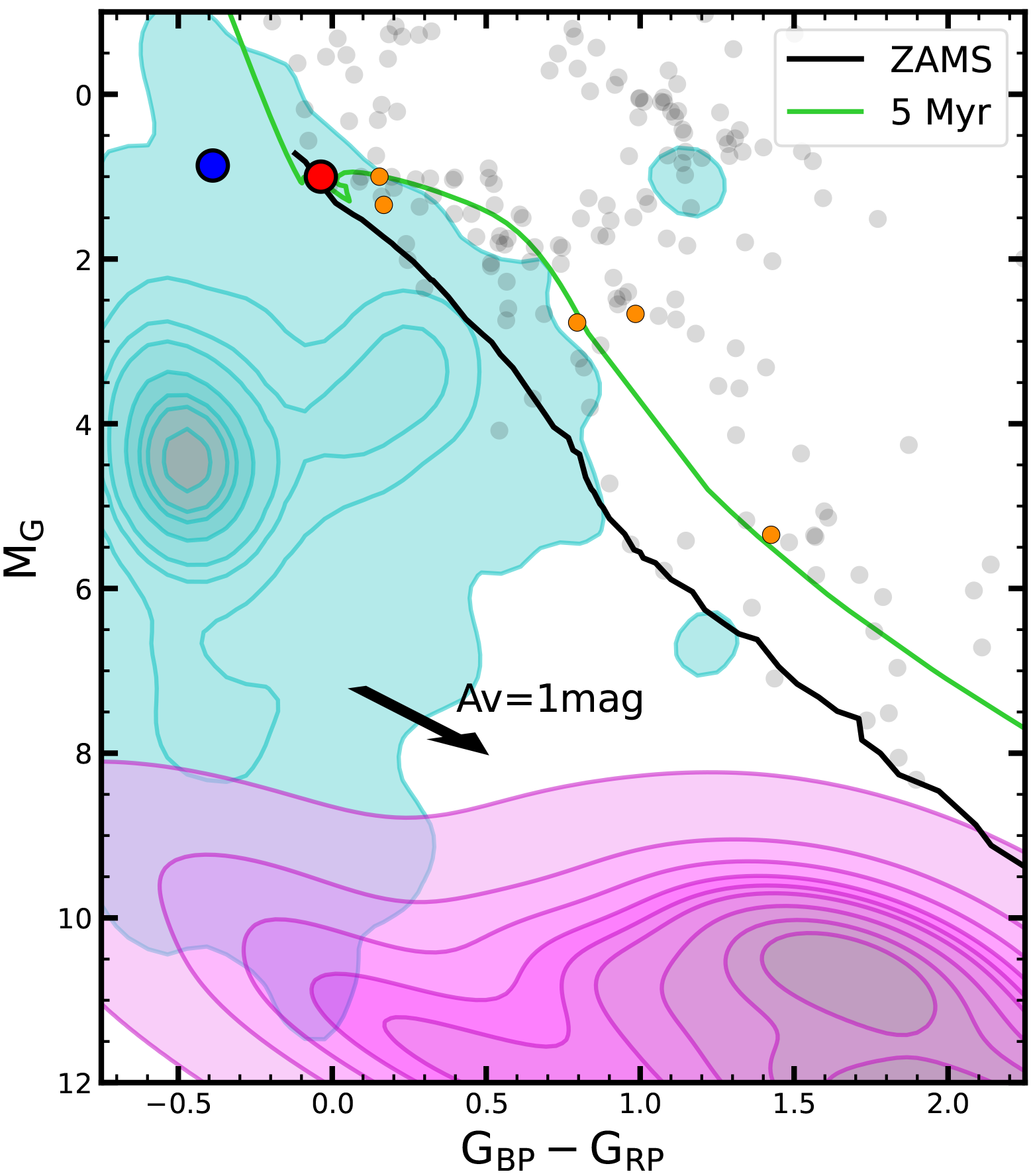}
    \caption{The Gaia CMD of GDR3 220\_28 (blue circle), aditionally representing a Herbig A0 type star of age 5 Myr (red circle) to observe the shift in its position. The figure also describes the distributions of 2332 sdB/O stars within 2.5 kpc of the solar neighbourhood (cyan, \citealp{2020yCat..36350193G}), and 112 WD+MS binaries (magenta, \citealp{2021MNRAS.506.5201R}). Grey circles represent neighboring sources, while orange circles represent sources identified as co-moving with GDR3 220\_28.}
    \label{CMD}
\end{center}
\end{figure}

In order to understand the relevant photometric characteristics of GDR3 220\_28, we examine its position in the CMD. Stars typically adhere to well-defined evolutionary sequences on the CMD \citep{2009AnApS...3..117V}. However, shifts from these expected positions can occur due to various factors, such as intrinsic variability or unresolved stellar companions.

The CMD for GDR3 220\_28 is plotted using the photometric magnitudes from Gaia DR3, which are corrected for extinction using $A_V$ derived from \citet{green}. The CMD is plotted between the absolute \textit{G} band magnitude $\mathrm{M_G}$ and the Gaia color (\textit{$G_{BP}-G_{RP}$}). $\mathrm{M_G}$ is calculated with the distance modulus relation. The CMD is shown in Figure \ref{CMD}. The CMD also features the ZAMS (\citealp{2013ApJS..208....9P}). The shift in the position of GDR3 220\_28 in the Gaia CMD is clearly observed, towards the left of the ZAMS, i.e., toward bluer colors. The grey circles represent sources from the initial sample of 50,953 objects that are located within a 25 pc volume of GDR3 220\_28. Their distribution in the CMD is broadly consistent with an age of 5 Myr, based on isochrones from Modules for Experiments in Stellar Astrophysics (MESA\footnote{MESA Isochrones \& Stellar Tracks Database: \url{http://waps.cfa.harvard.edu/MIST/}}, \citealp{choi2016mesa, dotter2016mesa}). The orange circles indicate sources identified as co-moving with GDR3 220\_28 from this neighboring sample using unsupervised k-means clustering in proper-motion space, suggesting a likely astrometric association with the source. These co-moving sources are also consistent with the 5 Myr isochrone. Accordingly, a representative age of $\sim5$ Myr is adopted for GDR3 220\_28. This is in agreement with literature recorded ages for the Cepheus Bubble of $\sim$7 Myr \citep{1998LNP...506..421P}. Variations in age of a few Myr do not significantly affect the analysis as an A0 type PMS star undergoes minimal changes along its evolutionary track in the CMD over a few Myr as it transitions into the main sequence phase \citep{1993ApJ...418..414P, 2016ApJS..222....8D}. Further, the position in the CMD (derived from MESA isochrones) of an A0 type PMS star of age 5 Myr is also indicated as a reference (The choice of spectral type is described in Sect. \ref{spec}).

However, such shifts in the CMD are not solely caused by physical systems; they can also arise due to erroneous measurements of distances and/or extinction. In order to rule out this possibility, we examine these measurements for the source. The measured parallaxes to the source in literature are entirely from the Gaia mission -- DR3 ($1.1275\ mas$) and earlier releases (DR2: $1.0664\ mas$). A variation of $\sim0.7\ mas$ in the Gaia DR3 parallax places the object back into the ZAMS. However, given Gaia's astrometric accuracy, this level of error is improbable, and therefore, we discredit the possibility that the shift in the CMD of the source is due to an error in the measured parallax. We further explore the potential for measurement error in determining the extinction, $A_V$, to the source. The query to the 3D dust reddening map by \citet{green} yields $A_V$ = 2.91. To verify its accuracy, we employ an alternative method which makes use of the diffuse interstellar band absorption strength to estimate extinction \citep{2011ApJ...727...33F, 2022ApJ...940..156C, 2023MNRAS.524.5166N}. Specifically, we use the method described by \citet{2022ApJ...940..156C}, which utilizes the absorption strength of the diffuse interstellar band at 5780\AA~to empirically estimate the dust reddening to any source. Across the observed epochs, the 5780~\AA\ DIB has a median EW of $0.47$~\AA, corresponding to an extinction of $A_V=2.99$ mag for GDR3 220\_28. The consistency between the spectral estimation of $A_V$ and that obtained from the 3D dust reddening map lends credibility to the value of $A_V$ used in this study ($A_V$ = 2.91).

\subsection{Spectroscopy}\label{spec}

In order to identify the spectral type of GDR3 220\_28, we fit the spectra of the source obtained from the HCT spectroscopic facility with the observed template spectra from the MILES spectral library \citep{2010MNRAS.404.1639V}. The template stars provided in the MILES library cover a wide range of effective temperatures (T$_{\text{eff}}$: 3000 – 40,000K) and surface gravity ($logg$: 0.2 – 5.5). In order to accurately estimate the spectral type, we match only the higher-order Balmer lines (H$\delta$, H$\epsilon$ and H$\zeta$) as emission mechanisms can fill in the lower-order Balmer lines. From the minimized chi-square, the best-fitting  MILES template is the A0-type star HD 31295. The literature parameters associated with with HD 31295 are $\log g = 4.08$ and $v\sin i = 130$ km/s. However, estimating accurate spectral type by template matching for low-resolution spectra is sub-optimal and one can expect deviations of up to 2 sub-classes in the case of A type stars \citep{2021MNRAS.501.5927A, 2021RAA....21..288S}.

\subsubsection{Multi-epoch spectroscopy}

\begin{figure}
\begin{center}
	\includegraphics[width=\columnwidth]{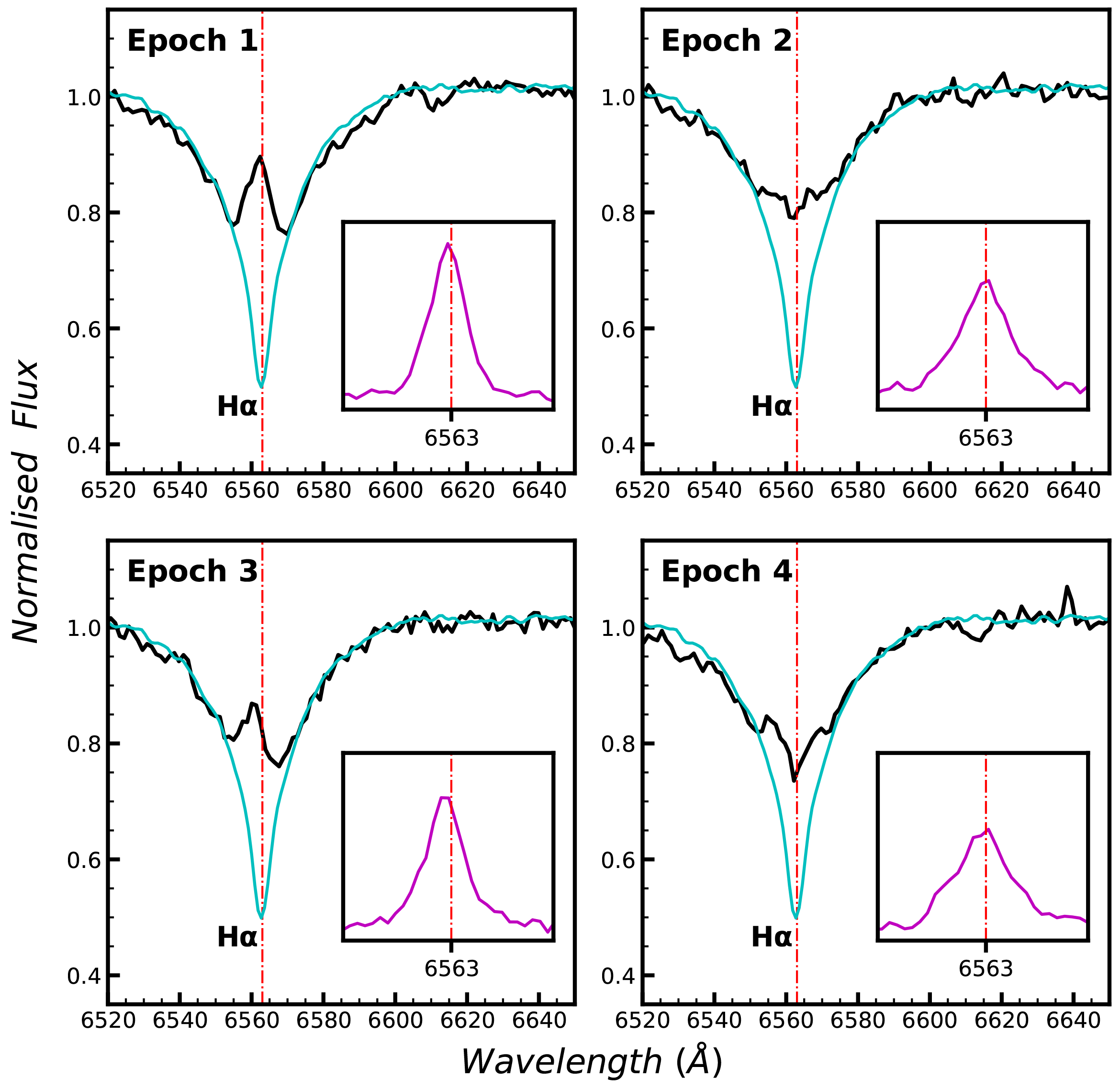}
    \caption{The H$\alpha$ line profiles for different epochs are displayed. The cyan spectrum represents the best fit template spectrum from the MILES library and the black spectra represents the observed spectrum at the mentioned epochs. The inset plots distinctly illustrate the emission component of the H$\alpha$ line. The red dashed line represents the rest wavelength of the H$\alpha$ line.}
    \label{Halpha}
\end{center}
\end{figure}

The variation in the profile of the H$\alpha$ line is studied over multiple epochs using the HCT spectra. All of the observed spectra feature an `emission in absorption' profile for the H$\alpha$ line, with net absorption. On measuring the EW of the H$\alpha$ line over the four epochs, we observe an increase in the absorption strength. The emission component in the H$\alpha$ line is isolated by subtracting the observed spectra at multiple epochs from the re-sampled MILES spectra. The EW of the H$\alpha$ emission component over the four epochs is measured and we observe decreasing EW measurements along the epochs. The line profiles for the H$\alpha$ line over the four epochs along with the isolated emission components are displayed in Figure \ref{Halpha}. Further, we also measure the full width at half maxima (FWHM) for the emission components (all of which are tabulated in Table \ref{Halphav}).

{\renewcommand{\arraystretch}{1}
\begin{table}
    \centering
    \caption{The H$\alpha$ line EW  and FWHM measured for the emission component over the four different epochs. The best fit MILES template spectrum has a measured H$\alpha$ EW of 11 \AA\ (in absorption).}
    \begin{tabular}{|c|c|c|c|}
    \hline
         &EW &EW of H$\alpha$&FWHM of H$\alpha$\\
         Epoch& of H$\alpha$&Em. Comp. &Em. Comp.\\
         &(\AA)&(\AA)&(\AA)\\
         \hline
         Epoch 1 &7.9&(-) 3.1&7.0\\
         (2023 Jan 16) &&&($\sim$320 km/s)\\
         \hline
         Epoch 2 &8.1&(-) 2.9&9.5\\
         (2023 Jan 21) &&&($\sim$430 km/s)\\
         \hline
         Epoch 3 &8.4&(-) 2.6&7.2\\
         (2023 Jan 31) &&&($\sim$330 km/s)\\
         \hline
         Epoch 4 &8.5&(-) 2.5&10.4\\
         (2023 Feb 8) &&&($\sim$480 km/s)\\
         \hline
    \end{tabular}
    \label{Halphav}
\end{table}}

The emission component of the H$\alpha$ line of GDR3 220\_28 is estimated to have EWs in the range 3.1 to 2.5 \AA, steadily decreasing over the 4 epochs. This nature of variability observed in the H$\alpha$ emission strength is most likely a consequence of instabilities in its circumstellar environment involving accretion \citep{cody2014csi}. This is further strengthened by the short term low amplitude variability in the WISE \textit{W1} and \textit{W2} photometric bands reported for the source in \citet{mei2023neowise}.

Another interesting observation is the variation seen in the measured FWHM of the emission components over the four epochs. At epochs 1 and 3, the FWHM is measured to be $\sim 7$ \AA~($\sim$320 km/s), and at epochs 2 and 4, the FWHM is measured to be $\sim 10$ \AA~($\sim$450 km/s). Furthermore, a consistent observation is the blueshift of the emission component by approximately 2 \AA~from the rest wavelength of the H$\alpha$ line at epochs 1 and 3. This recurring pattern between epochs implies a genuine phenomenon rather than an artifact resulting from spectral reduction. These variations, especially in the measured FWHM could be traces of radial-velocity shifts brought about due to a possible binary companion, with observed radial-velocity variations of upto 300 km/s in the case of sdB/O binaries \citep{2022A&A...661A.113G}, or through the contribution of H$\alpha$ emission periodically from a different region. However, due to the low resolution nature of the observed spectra, it is a perilous task to infer any further scientific insights from this observed variability seen in the FWHM measurements over the four epochs. To be able to conclusively derive the physical processes warrants high-resolution spectroscopic observations over multiple epochs.

\subsubsection{Blue Excess in the Gaia XP Spectrum}\label{gaia_blue}

The best-fit spectral classification of GDR3 220\_28 points to an A0-type star. Given the blue-ward position of the source in the Gaia CMD, we compare the Gaia DR3 BP/RP (XP) sampled spectrum of GDR3 220\_28 with that of HD 40573, a well-characterised A0-type optical spectrophotometric standard \citep{2016MNRAS.455.3864A}.

The Gaia XP spectra of both sources are corrected for extinction using the \citet{1989ApJ...345..245C} extinction law (with $R_V=3.1$). For HD 40573, we adopt $A_V=0.189$ mag \citep{2024A&A...691A..98K}. For comparison, the flux of HD 40573 is scaled from its distance of 188.19 pc \citep{2021AJ....161..147B} to that of of GDR3 220\_28. To account for any residual differences in the absolute scaling of the two stellar spectra, the HD 40573 spectrum is normalized to GDR3 220\_28 over the 8500--9500~\AA\ interval, where the contribution from a possible hot companion is expected to be small. The required scaling factor is 0.988, indicating that the two spectra agree within $\sim1\%$ in the normalization window.

\begin{figure}
\begin{center}
	\includegraphics[width=\columnwidth]{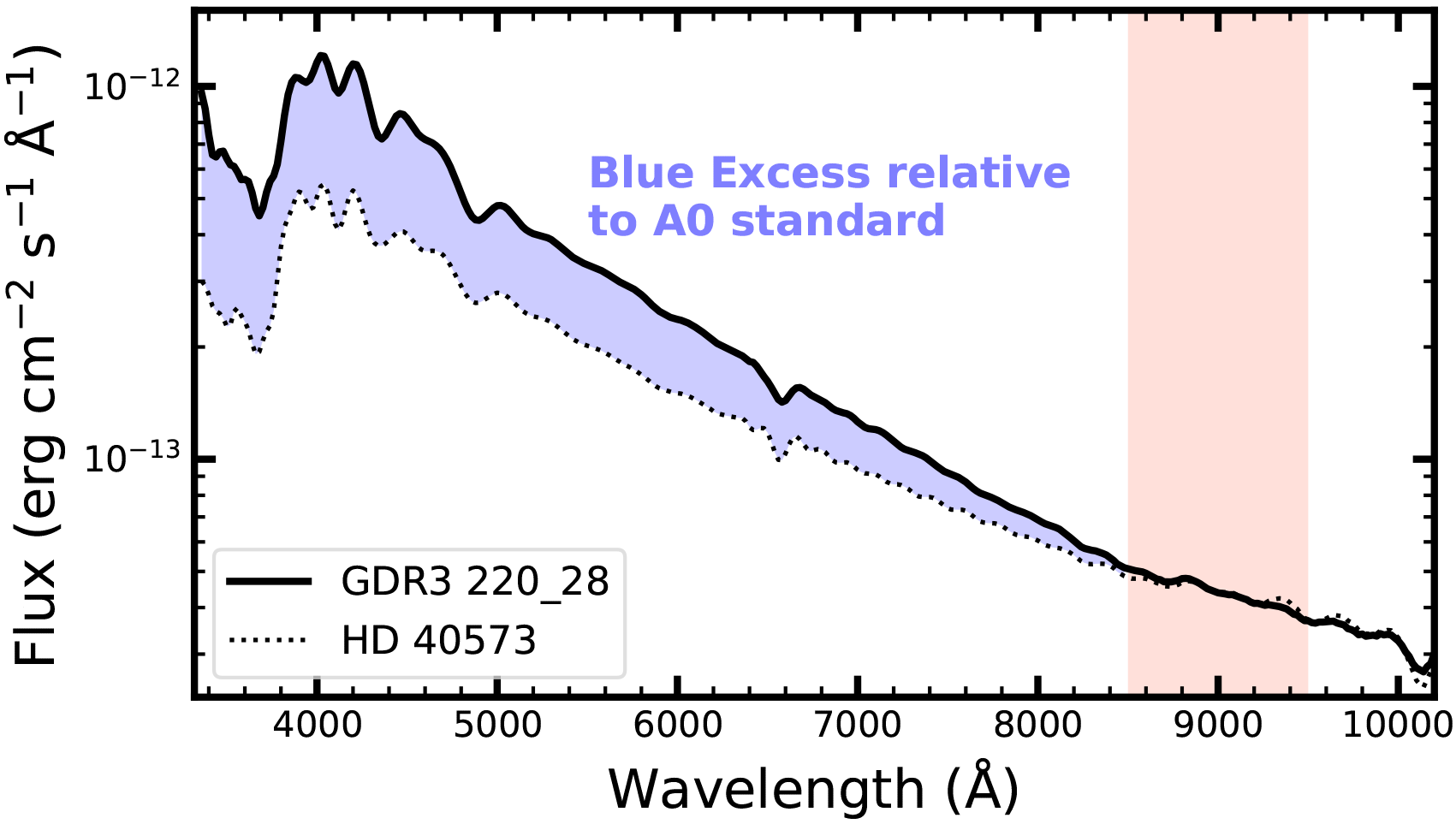}
    \caption{Extinction-corrected Gaia DR3 XP sampled spectrum of GDR3 220\_28 (black solid line) compared with that of the A-type spectrophotometric standard HD 40573 (black dotted line). The red shaded region shows the normalization window (8500--9500~\AA). The blue shaded region highlights the increasing excess of GDR3 220\_28 relative to HD 40573.}
    \label{GaiaXP}
\end{center}
\end{figure}

Towards bluer wavelengths, the spectrum of GDR3 220\_28 exhibits a systematic and increasingly strong excess relative to the A0 comparison spectrum (see Fig.~\ref{GaiaXP}). This consistent rise in the blue excess supports the presence of an additional hotter component contributing to the total flux of the system, providing an independent spectroscopic support for a hotter unresolved companion.

\subsection{Two-component SED Modeling}\label{sed_model}

\begin{figure}
\begin{center}
	\includegraphics[width=\columnwidth]{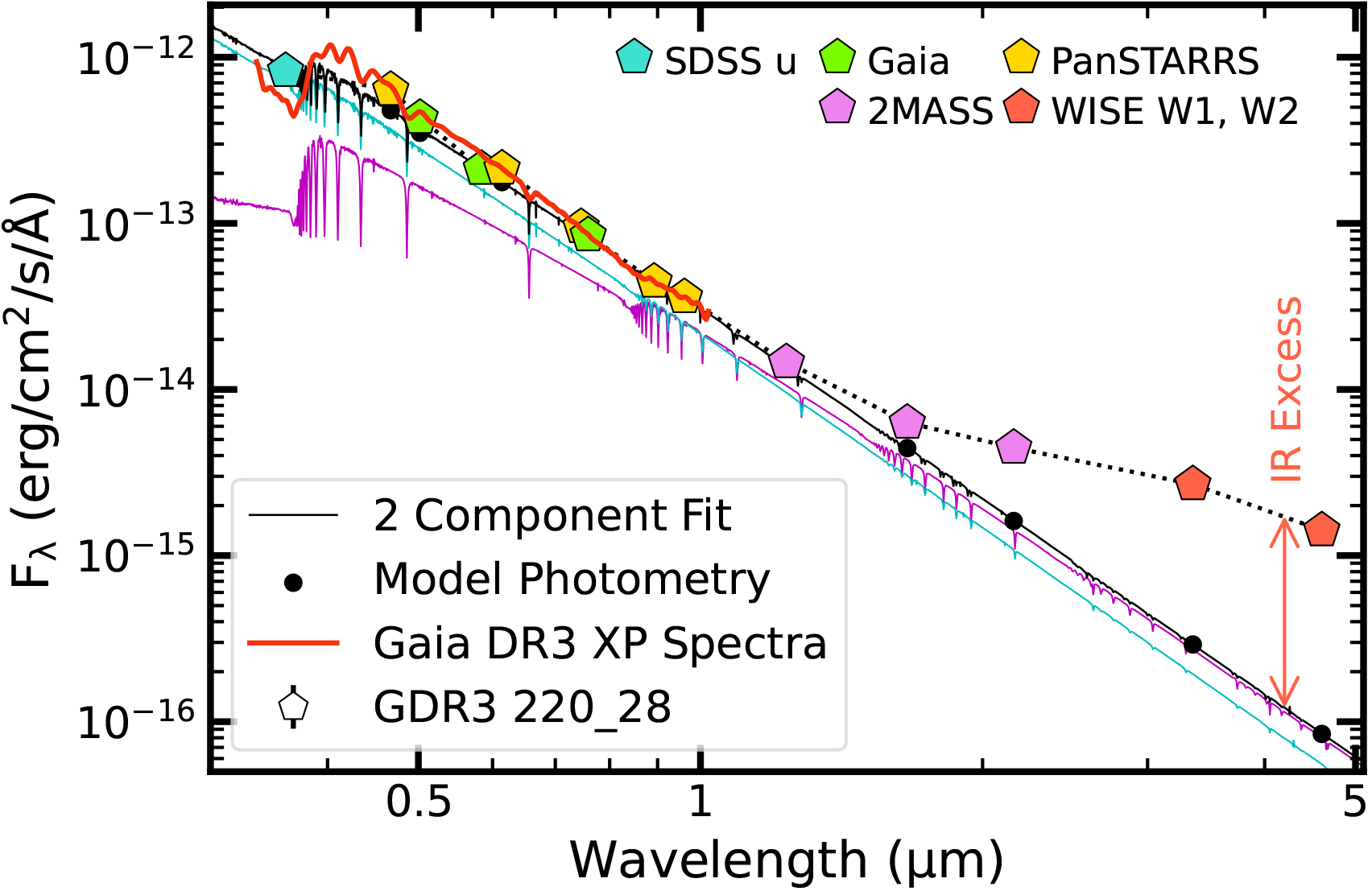}
    \caption{The SED of GDR3 220\_28. The photometry of the source in the various bands is given by the pentagons. The Gaia DR3 XP spectrum is also shown. The black curve shows the combined two-component spectrum. The individual components are shown in magenta ($T_{\rm eff}=9800$) and cyan ($T_{\rm eff}=27000$).}
    \label{SED}
\end{center}
\end{figure}

The blue excess observed in the Gaia XP spectrum motivates a two-component fit to the SED of GDR3 220\_28. The SED is plotted from available photometry in the SDSS \textit{u} band \citep{1996AJ....111.1748F}, the Gaia \textit{G}, \textit{$G_{BP}$} and \textit{$G_{RP}$} bands \citep{2023A&A...674A...1G}, the PanSTARRS \textit{g}, \textit{r}, \textit{i}, \textit{z} and \textit{y} bands \citep{2016arXiv161205560C}, the 2MASS \textit{J}, \textit{H} and \textit{K$_s$} bands \citep{skrutskie2006two}, and the WISE \textit{W1}, and \textit{W2} bands \citep{cutri2013explanatory, cutri2021vizier} (see Fig. \ref{SED}). Since GDR3 220\_28 is spectroscopically classified as A0 in Section \ref{spec}, one component is fixed at $T_{\rm eff}=9800$ K and $\log g=4.0$, while the second is allowed to vary. Model photometry is obtained from the BT-Settl AGSS2009 grid \citep{2003IAUS..211..325A, 2012RSPTA.370.2765A}, with dilution factors applied to the model fluxes to account for the relative contributions of the two components, and the combined model is fitted to the observed SED.

The best-fit second component has $T_{\rm eff}=27000$ K and $\log g=4.5$. It is important to note that the upper bound for surface gravity available in the BT-Settl grid for models with $T_{\rm eff}>10000$ K is $\log g=4.5$. As a result, the fitted value is regarded as a lower limit on the surface gravity of the hotter component. These parameters are consistent with hot subdwarf stars, specifically sdB stars \citep{2022A&A...661A.113G, 2026A&A...707A...6D}. The adopted/inferred stellar parameters of the two components from the analysis presented thus far are summarized in Table~\ref{params}.\\

{\renewcommand{\arraystretch}{1.5}
\begin{table}
    \centering
    \caption{Adopted/inferred stellar parameters of the two components of GDR3 220\_28.}
    \begin{tabular}{|c|c|c|}
    \hline
         Parameter&Component 1&Component 2\\
         \hline
         $T_{\rm eff}$ (K)&\hspace{-1cm}\makecell{$9800 \pm 600$\\{\footnotesize (Sect. \ref{spec})}}&\hspace{-1cm}\makecell{$27000\pm1000$\\{\footnotesize (Sect. \ref{sed_model})}}\\
         \hline
         $\log g$ (cgs)&\hspace{-1cm}\makecell{4.08\\{\footnotesize (Adopted from}\\{\footnotesize HD 31295)}}&\hspace{-1cm}\makecell{$\ge4.5$\\{\footnotesize (Sect. \ref{sed_model})}}\\
         \hline
         Radius ($R_{\odot}$)&\hspace{-1cm}\makecell{$2.03 \pm 0.02$\\{\footnotesize (Adopted from}\\{\footnotesize HD 40573)}}&\hspace{-1cm}\makecell{$\le0.64^{\dagger}$\\{\footnotesize (Using $\log g\ge4.5$)}}\\
         \hline
         Object Type &PMS Star&sdB candidate\\
         \hline
    \end{tabular}
    \label{params}
    \begin{minipage}{\linewidth}
    \footnotesize
    $^{\dagger}$The upper limit on the radius for Component 2 is estimated from $g=GM_{\star}/R_{\star}^2$, adopting the median sdB mass of $M_{\star}=0.471~M_{\odot}$ \citep{2012A&A...539A..12F}. For comparison, adopting $\log g=5.45$, characteristic of the
    cooler He-poor sdB population around $T_{\rm eff}\sim28000$ K
    \citep{2012MNRAS.427.2180N}, gives a radius of
    $R_{\star}\simeq0.21~R_{\odot}$.
    \end{minipage}
\end{table}}

The SED is also used to trace circumstellar dust and disk emission \citep{malfait1999iso, natta1993temperature}. In addition to the blue excess, GDR3 220\_28 shows clear near-IR excess beginning longward of the 2MASS \textit{H} band, as expected from its initial selection based on near- and mid-IR excess, indicating the presence of circumstellar dust.

\subsection{Comparison of GDR3 220\_28 with Analogous Stars in Literature}\label{similar}

\subsubsection{`Blue Stars with Excesses'}

PMS stars with shifted positions bluer than the PMS locus have been identified in literature as `blue stars with excesses' \citep{2010A&A...521A..18G, 2013MNRAS.435.3058D}. While most of these stars do not travel beyond the ZAMS in the CMD, there are rare instances where they shift even further beyond it \citep{1997AJ....113.1733H}. An explanation for `blue stars with excess' are that they experience strong accretion, characterized by a significant accretion continuum or `veiling' that can alter their optical colors \citep{1990ApJ...349..190H, 1997AJ....113.1733H, 2010A&A...521A..18G}. The critical mass accretion rate ($\dot M_{c}$) beyond which veiling can affect the observed optical colors of a PMS star is given by equation 10 in \citet{1990ApJ...349..190H}. For an A0 type PMS star, we estimate, $\dot M_{c} \sim 10^{-6}\ M_{\odot}yr^{-1}$. The average mass accretion rate ($\dot{M}_{acc}$) for GDR3 220\_28 is estimated using the procedure detailed in \citet{2019AJ....157..159A} and \cite{2023MNRAS.524.5166N}. We estimate $\dot{M}_{acc}$ for GDR3 220\_28 to be $6.97\pm1.88\times10^{-8}\ M_{\odot}yr^{-1}$. This estimate lies at the lower bounds of reported accretion rates for HAeBe stars ($10^{-6}$ to $10^{-9}\ M_{\odot}yr^{-1}$; \citealp{2006A&A...459..837G, wichittanakon}). This suggests that the source is nearing the end of its accretion phase. The estimated $\dot M_{acc}=6.97\pm1.88\times10^{-8}\ M_{\odot}yr^{-1}$ for GDR3 220\_28 is considerably lower than the required critical mass accretion rate. Thus, we dismiss the possibility of intense veiling causing the bluer colors observed for GDR3 220\_28.

Another explanation for `blue stars with excess' involves the presence of a disk that strongly scatters the stellar optical photons back into the line of sight, resulting in an excess of observed flux at optical wavelengths \citep{2010A&A...521A..18G}. The efficiency of this mechanism lies within the presence of small dust grains ($\le 1~\mu m$) in the disk, as they scatter optical photons more efficiently \citep{2001Sci...292.1686T}. However, considering the evolved state of the disk in GDR3 220\_28, we can conclude that dust grains of sizes $\le 1~\mu m$, required to scatter optical photons back into the line of sight, are scarce \citep{2000prpl.conf..533B, 2002ApJ...568.1008C, 2004ASPC..324...20N, 2004A&A...421.1075D}. Furthermore, for such a system, the disk would need to be inclined, obstructing light from the central star. Following the analysis presented in Section \ref{gaia_blue}, the observed flux of GDR3 220\_28 is in close agreement, within $\sim1\%$, with that of the A0 spectrophotometric standard star HD 40573 over the 8500--9500~\AA\ window. This close agreement leaves little evidence for additional extinction from an obscured disk configuration, making this scenario improbable for explaining its photometric characteristics.

\subsubsection{HAeBe stars}

From literature, we identify GDR3 220\_28 as a candidate HAeBe star from a catalog of 1361 HAeBe candidates \citep{2020A&A...638A..21V}. Consequently, we compare it with the rest of the catalog. This revealed 11 other HAeBe candidates which showed similar shifts to GDR3 220\_28 towards bluer colors. The brightest of these 11 HAeBe candidates (\textit{V} band magnitude $<$ 14), V* TZ Per and UCAC4 479-019568 have been reported as a Cataclysmic Binary \citep{2024ApJ...962..123S} and an sdB/O candidate \citep{2022ApJ...928...20B}, respectively. Further investigation of these candidates could potentially reveal additional insights into GDR3 220\_28 and similar classes of objects.

\subsection{Modeling an Unresolved Companion in the CMD}\label{binary}

We explore the possibility of an unresolved close binary companion due to the shift in its position in the Gaia CMD from the intended position of GDR3 220\_28. It is important to note that this analysis is made considering non-erroneous photometry in the Gaia bands (see Sect. \ref{phot}). The shift in color is toward the bluer region by a factor of 0.339 in the $G_{BP} - G_{RP}$ axis and the shift in the absolute magnitude $M_G$ is observed to be 0.113 toward the brighter end. We modeled this shift to account for a possible unresolved binary companion using the relation given below (Eq. \ref{mod}) which convolves the fluxes from two component sources into an unresolved total flux.

\begin{equation}
    m_{tot}=-2.5*log_{10}(A\times10^{-0.4m_1} + B\times10^{-0.4m_2}\vspace{0.5cm})
    \label{mod}
\end{equation}

\begin{figure}
\begin{center}
	\includegraphics[width=\columnwidth]{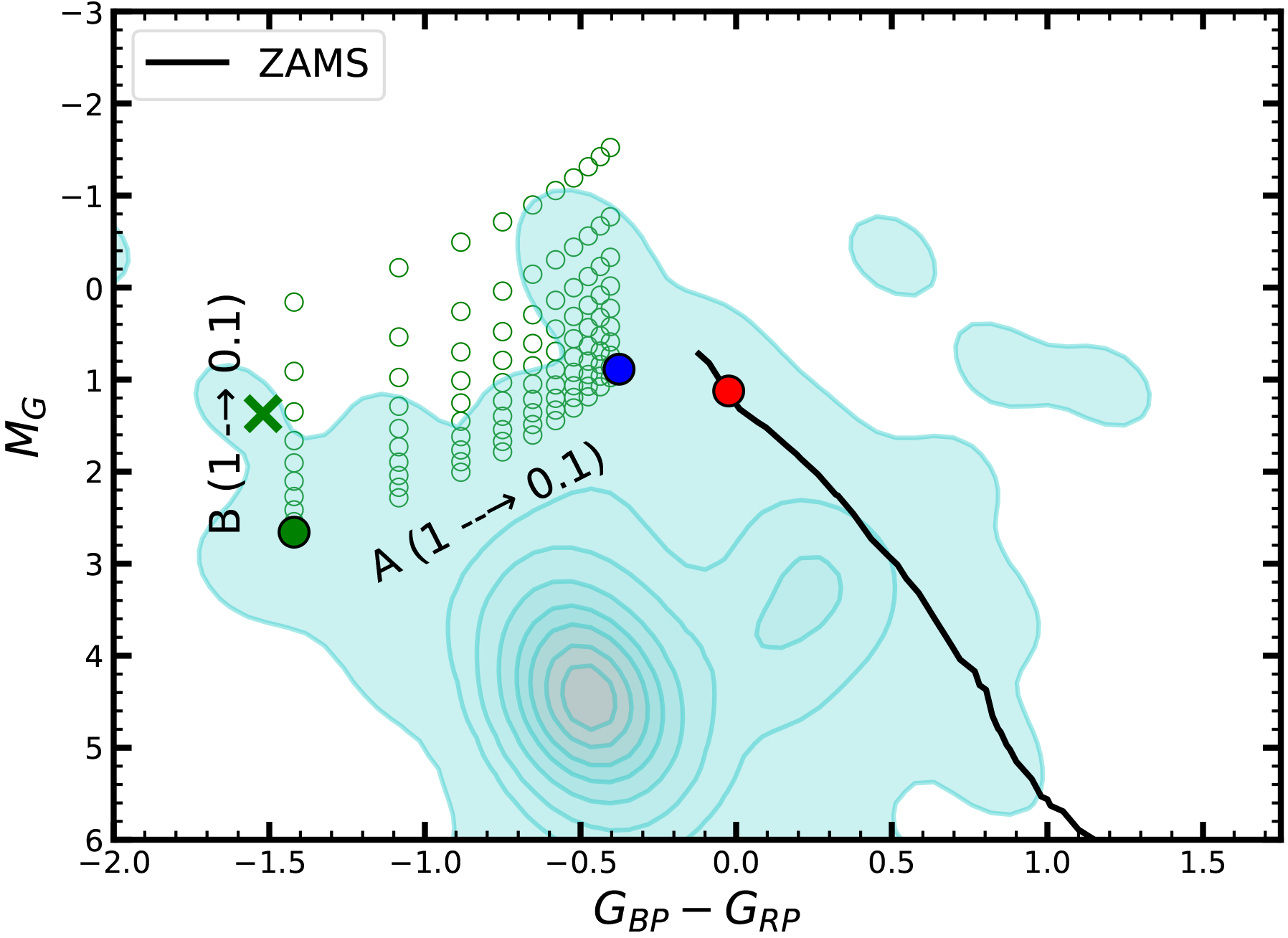}
    \caption{Gaia CMD of GDR3 220\_28. The green closed circle represents the position of a possible unresolved binary companion capable of producing the necessary shift seen in the CMD of GDR3 220\_28 (blue circle) when compared to an A0 type PMS star (red circle; \citealp{2013ApJS..208....9P}). The green open circles define the region where a binary component can exist by varying the dilution factors of the component fluxes. The green X marks the position of the companion independently derived from the Gaia XP spectral excess. The cyan contours describe a population of 1091 sdB/O stars that lie between 0.5 and 1.5 kpc of the solar neighbourhood \citep{2020yCat..36350193G}.}
    \label{CMDAna}
\end{center}
\end{figure}

Where $m_{tot},\ m_1$ (A0 type PMS star of age 5 Myr) $and\ m_2$ (possible unresolved companion) are the apparent magnitudes of the unresolved system and the 2 resolved components, respectively. Further, A and B are ‘dilution constants' which ascribe the individual contributions of the 2 components ($m_1\ and\ m_2$) to the total unresolved flux \citep{bayo2008vosa, gomez2022hot}. The position in the CMD modelled for the possible unresolved companion is shown in Figure \ref{CMDAna}.

Further, as an independent check on the position of the unresolved companion in the CMD, we follow up on the blue excess identified in the Gaia XP spectrum of GDR3 220\_28 (Sect. \ref{gaia_blue}). The photometry of the unresolved companion is estimated by integrating the residual flux over the Gaia \textit{G}, \textit{$G_{BP}$}, and \textit{$G_{RP}$} passbands \citep{2021A&A...649A...3R}. This residual flux is calculated by subtracting the extinction-corrected XP spectrum of HD 40573 from that of GDR3 220\_28, which is then re-extincted assuming the same line-of-sight extinction as GDR3 220\_28. The resulting position is shown in Figure \ref{CMDAna}.

The region occupied in the CMD by the unknown binary companion is one sparingly populated ($\sim$2\%) by sdB/O star populations (see Fig. \ref{CMDAna}). Although they are primarily known to exist in binary systems with degenerate objects, such as white dwarfs, sdB/O stars have been discovered in binary systems with non-degenerate objects such as early-type main sequence stars \citep{2015MNRAS.450.3514K, 2016PASP..128h2001H, 2021AJ....161..248W}. Further, \citet{2018MNRAS.475.2480P} describes binary systems of main sequence stars with sdB/O companions to explain `sdA stars,' showing H-rich spectra. The two-component SED analysis presented in Section~\ref{sed_model} yields $T_{\rm{eff}} = 27000~\mathrm{K}$ for the second component, consistent with sdB stars. Binary population synthesis studies reveal the birthrates of sdB stars in the `common envelope' channel in a short-period binary (orbital periods ranging from 0.1 to 10 days) with a main-sequence companion are $\approx 0.7\times10^{-2}~yr^{-1}$ \citep{2002MNRAS.336..449H, 2003MNRAS.341..669H, 2008ASPC..392...15P}.

\begin{figure}
\begin{center}
	\includegraphics[width=\columnwidth]{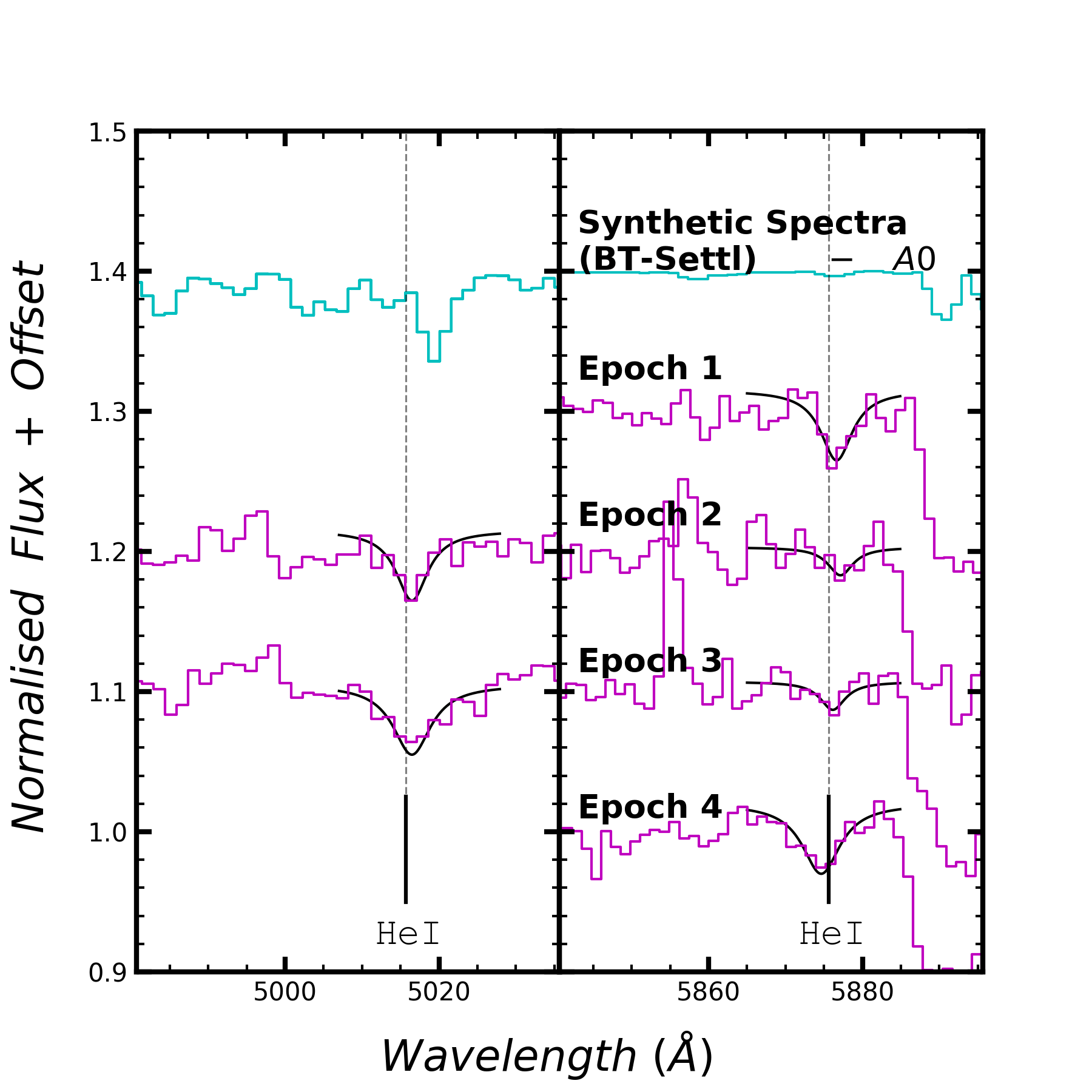}
    \caption{The He{\sc i} (5016 \AA~and 5876 \AA) line profiles at the various epochs are shown. The figure additionally shows an A0 type synthetic stellar spectrum (BT-Settl; \citealp{2003IAUS..211..325A}) in cyan, resampled and broadened to match the resolution and $vsini$ of the observed spectra. Gaussian profiles are overlaid against each observation for representation. Note: The He{\sc i} 5016 \AA~line is observed with the Gr7 grating (observations using Gr7 is made only at 2 out of the total 4 epochs).}
    \label{HeVar}
\end{center}
\end{figure}

The presence of a hot companion (such as an sdB star) in a close binary system typically results in hot spectral lines in the observed spectrum and notable ultraviolet (UV) excesses as they contribute relatively more flux at UV wavelengths \citep{2017ApJ...843...60W, 2018OAst...27...35H}. This, along with the fact that we don't expect to see the presence of hot spectral lines in the spectra of an A0-type star \citep{2009ssc..book.....G}, ensures that any such lines detected are likely contributed by a hot companion. We comb through the optical spectra of GDR3 220\_28 observed from HCT to look for hot spectral lines (such as He{\sc i}, He{\sc ii}, and other high energy lines) that could further strengthen the presence of a hot companion. However, the strength of helium lines in sdB/O stars depend strongly on atmospheric He abundance, which broadly increases with $T_{\rm eff}$ \citep{2012MNRAS.427.2180N}. At $T_{\rm eff}\sim27000$ K, sdBs can have very low He abundances, and consequently weak or undetectable He{\sc i} features \citep{2012MNRAS.427.2180N, 2026A&A...707A...6D}. Using the empirical He abundance relations for He-rich and He-poor sdB sequences given by equations 2 and 4 of \citet{2012MNRAS.427.2180N}, $T_{\rm eff}=27000$ K corresponds to very low He abundances of [He/H]$_* \simeq -2.59$ and $-3.78$, respectively. The relatively low resolving power of the HCT spectra ($R\sim1200-2000$) further limits their identification. Despite these limitations, we detect 2 He{\sc i} lines in the optical spectra of GDR3 220\_28, with their profiles showing variability across different epochs (see Fig. \ref{HeVar}). However, the low resolution of the available optical spectra limits our ability to accurately quantify these variations.

The analysis carried out in this subsection, describing a binary system, effectively explains the observed shift in its optical photometry. Furthermore, this scenario successfully clarifies other features seen in its spectra and photometry. A binary companion can explain the variations noted in the FWHM values recorded for the H$\alpha$ emission over the 4 epochs. With respect to the IR excess observed in the SED of GDR3 220\_28, while a circumbinary disk can substantiate the presence of mid-IR excesses in the SED of binary systems, the presence of near-IR excess requires further consideration. However, this could possibly be attributed to the PMS nature of GDR3 220\_28, where a disk characterised by a puffed inner rim surrounding the star likely contributes to the near-IR emission \citep{2006A&A...459..837G, 2014prpl.conf..339T}.

\subsection{Light curve Analysis}

The possibility of a binary companion, as discussed in Section \ref{binary}, along with the observed spectral indications, warrants a check for periodic variability attributed to the source. 

\begin{figure*}
\begin{center}
    \includegraphics[width=2\columnwidth]{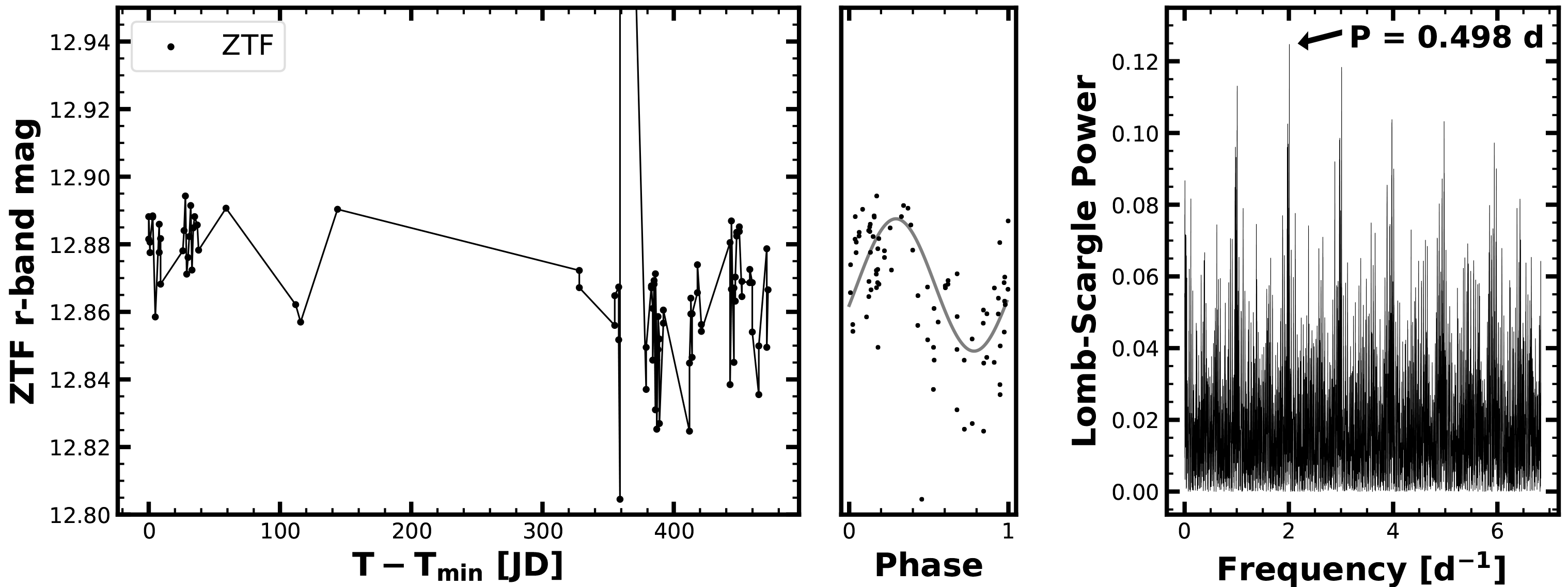}
    \caption{(Left) ZTF light curve of GDR3 220\_28 and (Middle) the phase folded light curve at a period of 0.498 days. (Right) The Lomb-Scargle periodogram, showing a prominent peak at 2.008 $d^{-1}$.}
    \label{LC}
\end{center}
\end{figure*}

We queried the TESS archives \citep{2015JATIS...1a4003R} for GDR3 220\_28, which returned time series data that exhibits a sinusoidal variability. Using the Python package {\it Lightkurve}\footnote{Lightkurve documentation: \url{https://docs.lightkurve.org/}}, we derived a periodicity of 0.4857 days. However, when we performed a query of nearby sources to rule out any effects of contamination due to the low spatial resolution of TESS ($\sim21\arcsec$/px), we found a nearby bright source (\textit{V} band magnitude $\sim11$) at a separation of $\sim13\arcsec$, TIC 337831760, with identical time series data. This confirms that the TESS time series data of GDR3 220\_28 may be contaminated by the nearby source, TIC 337831760, or vice versa. We therefore do not use the TESS photometry in the subsequent analysis.

To avoid any contamination from nearby sources, we identified light curves for GDR3 220\_28 in the ZTF \citep{2019PASP..131a8002B, 2019PASP..131a8003M} archive, due to its higher spatial resolution. ZTF has a pixel scale of $\sim1^{\prime\prime}/\rm{px}$ and a median delivered image quality of about $2^{\prime\prime}.0$ FWHM in the $r$ band \citep{2019PASP..131a8002B}. To explore any periodicity associated with the source, we analyzed the ZTF light curves using the Lomb-Scargle periodogram \citep{2015ApJ...812...18V, 2018ApJS..236...16V}. Our analysis revealed a peak at 2.008 $d^{-1}$, corresponding to a period of 0.498 days, as shown in the Lomb-Scargle periodogram in Figure \ref{LC}. The figure also shows the ZTF light curve and the phase-folded light curve at the identified period, together with a sinusoidal fit. If associated with the binary system, this tentative period would be consistent with the short orbital periods reported for sdB/O binaries. More than half of the known sdB/O stars have been identified to exist in detached close binary systems of periods ranging from hours to days with separations of only a few stellar radii \citep{maxted2001binary, napiwotzki2004close, copperwheat2011sdss}.

\section{Summary}\label{summary}

In the work presented in this paper, we have undertaken a comprehensive analysis of an intriguing source within the Cepheus SFR. Our investigation has delved into various aspects, including the source's astrometry, photometry, and spectroscopic variability, in an effort to unravel the factors contributing to its unique characteristics. This pilot study emphasizes the potential significance of ongoing research on these unconventional objects in a broader context. Some of the results and observations made of the source are listed:

\begin{itemize}
    \item GDR3 220\_28 lies significantly blueward of the expected position of an A0-type PMS star in the Gaia CMD. The surrounding and co-moving sources are broadly consistent with an age of $\sim5$ Myr.
    \item We identified the spectral type of GDR3 220\_28 to be A0. Further, the HCT spectra show variable H$\alpha$ emission across multiple epochs. The observed FWHM variations may indicate a possible binary companion, although higher-resolution spectroscopy is needed for confirmation.
    \item The extinction-corrected Gaia XP spectrum shows a systematic blue excess relative to the A0 spectrophotometric standard HD 40573, providing independent evidence for the contribution of a hot companion.
    \item Two-component SED modelling, with the A0 component fixed at $T_{\rm eff}=9800$ K, yields a second hot component with $T_{\rm eff}\sim27000$ K and $\log g\gtrsim4.5$, placing it in the sdB/O regime, particularly an sdB candidate. The SED also shows a near-IR excess indicative of circumstellar dust.
    \item Alternative explanations for the blueward CMD shift, including strong accretion veiling and disk-scattered optical emission, are unlikely to fully account for the observed properties of the source.
    \item Our exploration of an unresolved close binary companion successfully modeled the shift observed in the CMD. Weak He{\sc i} features are seen in the HCT spectra while hotter spectral lines (such as He{\sc ii}) are absent, suggesting that the companion is likely an sdB star.
    \item The higher spatial-resolution ZTF light curve shows tentative evidence for a periodic variability of $0.498$ days, following a periodogram analysis. The corresponding phase-folded light curve shows a possible coherent variability. This is consistent with the short orbital periods commonly observed in close sdB/O binaries.
\end{itemize}

In conclusion, this comprehensive analysis has provided valuable insights into the nature of GDR3 220\_28. Our analysis supports the nature of GDR3 220\_28 as a binary system consisting of an A0-type PMS star ($T_{\rm eff}=9800$ K, $\log g=4.08$) and an sdB companion ($T_{\rm eff}=27000$ K, $\log g\geq4.5$). The rarity of such configurations highlights their importance in stellar evolution studies, and such analyses help address the observed under-representation of such systems relative to theoretical predictions from stellar population models. This work highlights the importance of multi-wavelength and multi-epoch observations in the study of such sources that lie shifted in their color-magnitude diagrams and serves as a foundation for future investigations into this intriguing class of astronomical objects. We intend to undertake a comprehensive global survey of similar objects, with these shifts serving as indicators pointing towards the existence of potential binary systems with elusive companions.

\begin{acknowledgments}
\nolinenumbers
The authors sincerely thank the anonymous reviewer for the careful and thorough evaluation and constructive suggestions, which significantly improved the focus and scientific quality of the manuscript. We extend sincere appreciation to Dr R. Arun, A. Devaraj and A.V. Sidharth for insightful discussions, which helped in shaping the ideas presented in this research. We are grateful to the Centre for Research, CHRIST (Deemed to be University), Bangalore, for the seed money research grants extended to carry out the current project (SMSS-2220 12/2022, SMSS-2335 11/2023). We would like to acknowledge and thank the Anusandhan National Research Foundation (ANRF), a statutory body of the Department of Science and Technology (DST), Government of India, for funding our research under Grant Number CRG/2023/005271. We also acknowledge facility support from the Department of Science and Technology (DST), Government of India, under the FIST programme (SR/FST/PS-I/2022/208). We thank the SIMBAD database and the online VizieR library service for helping us with the literature survey and obtaining relevant data. This work has made use of data from the European Space Agency mission {\it Gaia} (\url{https://www.cosmos.esa.int/Gaia}), processed by the {\it Gaia} Data Processing and Analysis Consortium (DPAC, \url{https://www.cosmos.esa.int/web/Gaia/dpac/consortium}). This publication makes use of data products from the Wide-field Infrared Survey Explorer and the Two Micron All Sky Survey. This research has made use of `Aladin sky atlas' developed at CDS, Strasbourg Observatory, France.
\end{acknowledgments}

\bibliography{Ref}{}
\bibliographystyle{aasjournal}

\end{document}